\documentclass[aps,twocolumn,superscriptaddress,showpacs,amssymb,prb,floatfix,preprintnumbers]{revtex4-1}

\usepackage{amsmath}
\usepackage{graphicx,color}
\usepackage[caption=false]{subfig}

\newcommand{\slrrtext}  {spin-lattice relaxation rate }
\newcommand{\slrr}      {$T_1^{-1}$\ }
\usepackage[dvipdfm]{hyperref}
\hypersetup{colorlinks=true,linkcolor=blue,
  implicit=true,breaklinks=true,pagebackref=true,backref=true,
  bookmarks=true,bookmarksnumbered=true,hyperfootnotes=true,debug=true,
  naturalnames=false,citecolor=blue,pdfview=FitH,pdfstartview=FitH,hyperindex=true}

\begin{document}

%\preprint{}

\title{Nuclear magnetic resonance as a probe of electronic states of Bi$_{2}$Se$_{3}$}


\author{D. M. Nisson}
\author{A. P. Dioguardi}
\author{P. Klavins}
\author{C. H. Lin}
\author{K. Shirer}
\author{A. Shockley}
\author{J. Crocker}
\author{N. J. Curro}
%\thanks{}
\affiliation{Department of Physics, University of California, Davis, CA 95616, USA}

\date{\today}

\begin{abstract}
We present magnetotransport and $^{209}$Bi nuclear magnetic resonance (NMR)  data on a series of single crystals of Bi$_{2}$Se$_{3}$, Bi$_2$Te$_2$Se and Cu$_x$Bi$_2$Se$_3$ with varying carrier concentrations. The Knight shift of the bulk nuclei is strongly correlated with the carrier concentration via a hyperfine coupling of 27 $\mu$eV, which may have important consequences for scattering of the protected surface states.  Surprisingly we find that the NMR linewidths and the spin lattice relaxation rate appear to be dominated by the presence of localized spins, which may be related to the presence of Se vacancies.
\end{abstract}


\maketitle

\section{Introduction}
Topological insulators are a novel form of condensed matter in which the bulk is electrically insulating but the surface electronic states remain gapless and conducting. \cite{KaneTIreviewRMP} This unusual situation can emerge in materials with large
spin-orbit couplings such that the topology of the bulk
band structure differs from that of the surrounding material.\cite{zhang2009} At the boundaries of the topological material the surface electron states exhibit a Dirac dispersion, and the number of such Dirac points in $\mathbf{k}$-space is odd. Furthermore,  the strong spin orbit coupling locks the electron spin direction to the momentum giving rise to chiral states.\cite{HasanQuantumSpinHallNature2008}   These chiral surface states are protected, meaning that perturbations such as defects and disorder on the surface will not alter their character.  More importantly electron backscattering is strongly suppressed, and may enable dissipationless polarized spin currents which could prove useful in spintronics applications.\cite{moore2010}

In 2009 the first of the bismuth chalcogenide family of materials, Bi$_2$Se$_3$, was discovered to exhibit surface states with the single Dirac cone characteristic of a 3D topological insulator.\cite{xia2009}   These compounds have attracted significant attention because large single crystals can be grown and clean surfaces can be exposed by cleaving, which enables detailed studies of the electronic dispersion via angle resolved photoemission spectroscopy (ARPES).\cite{KaneTIreviewRMP}  Bi$_2$Se$_3$ has a rhombohedral crystal structure consisting of hexagonal stacked planes of Bi and Se  (see Fig.~\ref{xrays}). The planes are organized into quintuple layers, and each layer has Se atoms on top and bottom.   In principle this material should be an intrinsic semiconductor with a gap of 350 meV; however, there is an inherent chemical tendency for Se atoms to be missing from the lattice.\cite{hyde1974}  These vacancy sites act as donors, raising the chemical potential out of the insulating regime and forming a degenerate semiconductor. In nearly all cases the vacancies place the Fermi level well in the conduction band, resulting in metallic  behavior. As a result the exotic transport properties of the surface states are often masked by the parallel bulk conduction channel.  \cite{butch2010}  Although several techniques have emerged to dope these materials so that the bulk states become insulating and the surface states are tuned to the Dirac point, relatively little is known about the microscopic electronic response of both the bulk and surface states.\cite{AndoBi2Se3Tuning,HorCuDopingReference,HorCaDopingReference} In order to investigate the bulk states in more detail we have conducted $^{209}$Bi nuclear magnetic resonance (NMR) and magnetotransport studies of Bi$_2$Se$_3$ grown in excess Se, Bi$_2$Te$_2$Se, and Cu doped Cu$_x$Bi$_2$Se$_3$. We find clear evidence of correlations between the NMR response and the carrier concentration in the Bi$_2$Se$_3$ samples.

Despite extensive research on the bismuth chalcogenides, there have been relatively few studies of the NMR response of this material.  $^{209}$Bi spectra and relaxation rates were reported in bulk Bi$_2$Se$_3$,\cite{young2012} and $^{125}$Te NMR has  been reported in nanoscale powders of Bi$_2$Te$_3$.\cite{LouiePRL2013} Depending on the surface to volume ratio, nuclei at the surface may  have sufficient spectral weight to contribute to the NMR signal, and in the former study the spectra revealed features consistent with surface nuclei.\cite{AnsermetSlichter} Moreover, the spin-lattice relaxation rate ($T_1^{-1}$) of these surface nuclei was enhanced compared to the bulk. The authors concluded that this enhancement was consistent with metallic protected surface states. The \slrrtext of the nuclei may be driven by spin-flip scattering between the electron and nuclear spins via the hyperfine interaction.  However, it is unclear whether \slrr is enhanced or suppressed by the exotic surface states, or whether the hyperfine interaction would have sufficient magnitude.  In order to understand the interaction of nuclei with the topological states  it is essential to understand the relaxation rates and hyperfine couplings of both the bulk and surface states.

The bismuth chalcogenides are well suited for NMR because Bi, Se and Te all have NMR active isotopes: $^{209}$Bi ($I=\frac{9}{2}$, 100\% abundant),
$^{77}$Se ($I=\frac{1}{2}$, 7.63\% abundant) and $^{125}$Te ($I=\frac{1}{2}$, 7.07\% abundant).\cite{rosman1998commission} $^{209}$Bi is particularly useful because not only is the signal intensity larger than that of $^{77}$Se, but it is also sensitive to the local charge environment via the quadrupolar interaction.  The nuclear spin Hamiltonian for $^{209}$Bi is given by:
\begin{equation}
\mathcal{H} = \gamma\hbar\hat{I}_zH_0 + \frac{h\nu_{cc}}{6}[3\hat{I}_z^2-\hat{I}^2 - \eta(\hat{I}_x^2 - \hat{I}_y^2)] + \mathcal{H}_{\rm hf},
\label{eqn:hamiltonian}
\end{equation}
where $\gamma=0.6842$ kHz/G is the gyromagnetic ratio, $H_0 = 9$ T  is the external field (except for sample \#1 and the Bi$_2$Te$_2$Se in which case $H_0=11.724$ T), $\hat{I}_{\alpha}$ are the nuclear spin operators, $\nu_{cc}$ is the component of the electric field gradient (EFG) tensor along the $c$-direction, $\eta$ is the asymmetry parameter of the EFG tensor, and $\mathcal{H}_{\rm hf}$ is the hyperfine interaction between the Bi nuclear spins and the electron spins. \cite{CPSbook}
Since there is a single Bi site in the unit cell with axial symmetry, $\eta$ is zero, and the spectrum consists of a set of nine equally spaced resonances separated by $\nu_{cc}$.  The hyperfine coupling can be written as $\mathcal{H}_{\rm hf} = \gamma \hbar g \mu_B A \mathbf{I} \cdot \mathbf{S}$, where $\mathbf{S}$ is the electron spin (of either the bulk carriers or of the surface state electrons) with g-factor $g$, and $A$ is the direct contact hyperfine interaction. The hyperfine coupling gives rise to Knight shift $K=K_0+A\chi$, where $\chi$ is the bulk magnetic susceptibility and $K_0$ is the temperature independent orbital shift.  For a degenerate semiconductor or metallic system, the susceptibility is given by $\chi= g^2\mu_B^2N(0)$, where $N(0)$ is the density of states at the Fermi level. The nuclear spin Hamiltonian (Eq. \ref{eqn:hamiltonian}) gives rise to nine resonances of the Bi nuclei in Bi$_2$Se$_3$ at frequencies:
 \begin{equation}
\omega_n = \gamma H_0 (1+K) + n\nu_{cc},
\label{eqn:freqs}
\end{equation}
 where  $n= -4, -3, \cdots +4$.

\section{Experimental Methods}

\begin{figure}
\includegraphics[width=\linewidth]{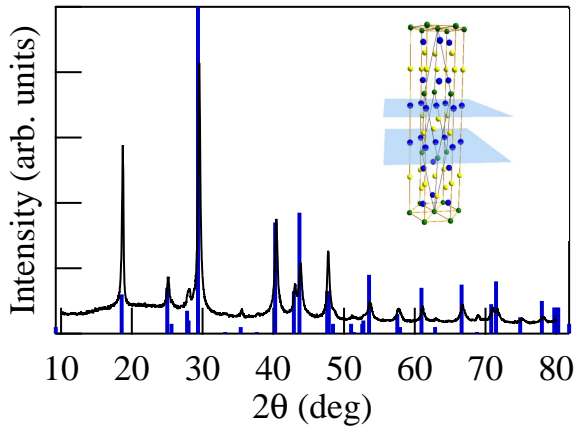}%
  \caption{\label{xrays}A representative powder X-ray diffraction
  	pattern of Bi$_2$Se$_3$ (black line) compared with  theoretical prediction
	(blue) Inset: Crystal structure of Bi$_2$Se$_3$. Bi
	(blue) and Se (green and yellow) atoms form quintuple layers.}
\end{figure}

\begingroup
\squeezetable
\begin{table*}[htbp]
  \centering
  \caption{\label{tab:samples}Measured Bi$_2$Se$_3$ samples, and their growth methods and properties at 10 K. The atomic percentage of Se for a stoichiometric Bi$_2$Se$_3$ mixture is 60\%.}
    \begin{ruledtabular}
    \begin{tabular}{cccrrrrrrrr}
    Sample & Batch & Growth mechanism & \% Se & $n$ (cm$^{-3}$) & $\mu \text{(V}^2 \text{m}^{-1} \text{s}^{-1}\text{)}$ & $f_{\text{SdH}}$  (T) & K  (\%) & $\Delta \omega_{\text{ctr}}$ (kHz) & $\Delta \nu_{\text{Q}}$ (kHz) & $\nu_{cc}$ (kHz)\\
    \hline
    \#1   &  A   & Ampoule anneal & 60 & N/A   & N/A   & N/A   & $0.647 \pm 0.002$ & $38 \pm 3$ & $25 \pm 3$ & $158.3 \pm 0.6$ \\
    \#2   & B   & Bridgman & 60 & $1.58 \times 10^{19}$ & 0.540 & $127 \pm 6$ & $0.619 \pm 0.001$ & $30 \pm 2$ & $30 \pm 2 $ & $160.5 \pm 0.5$\\
    \#3   & C   & Bridgman & 65 & N/A   & N/A   & $76 \pm 4$ & $0.51 \pm 0.01$ & $86 \pm 4$ & N/A & $168 \pm 2$\\
    \#4   & C   & Bridgman & 65 & $6.38 \times 10^{17}$ & 1.049 & $38 \pm 4$ & $0.356 \pm 0.003$ & $80 \pm 10$ & N/A & $159.9 \pm 0.06$\\
    \#5   & D   & Bridgman & 62.5 & $3.46 \times 10^{18}$ & 0.249 & N/A   & $0.441 \pm 0.004$ & $56 \pm 5$ & $11 \pm 16$ & $166 \pm 2$\\
    \#6   & E   & Bridgman & 60 & $1.35 \times 10^{19}$ & 0.108 & N/A   & $0.667 \pm 0.001$ & $37 \pm 2$ & $43 \pm 1$ & $150.5 \pm 0.5$\\
    \#7 & B & Bridgman & 60 & $9.41 \times 10^{18}$ & 0.645 & N/A & N/A & N/A\\
    \#8 & B & Bridgman & 60 & N/A & N/A & N/A & N/A & N/A\\
    \#9 & B & Bridgman & 60 & $1.91 \times 10^{19}$ & 0.680 & N/A & N/A & N/A\\
    \end{tabular}%
    \end{ruledtabular}%
\end{table*}%
\endgroup%

Several different single crystals of Bi$_2$Se$_3$ were prepared by the
Bridgman method from varying mixtures of elemental Bi and
Se in evacuated quartz ampoules.\cite{1925} A vertical Bridgman furnace was set up
%using a 1/140 rpm motor and gearbox
to lower an ampoule through a
temperature gradient of about $9\,^{\circ}\mathrm{C}$/cm around the
melting point of $710\,^{\circ}\mathrm{C}$ at a rate of 2 mm/hr.
%The temperatures of the gradient ranged from $750\,^{\circ}\mathrm{C}$ down to $650\,^{\circ}\mathrm{C}$.
Stoichiometric mixtures of
the elements were pre-melted and homogenized in the ampoules at
$800\,^{\circ}\mathrm{C}$ for 12 hours and then furnace cooled. The
ampoule was then placed into the
Bridgman furnace and lowered through the gradient. Samples that were
prepared with the initial mixture having a nominal stoichiometry of
Bi$_2$Se$_3$ formed an ingot with large columnar single crystals of a
preferred orientation. However, samples grown with an excess of Se
formed polycrystalline ingots, from which only a few single crystal
samples could be obtained.  We found that samples grown simply by annealing in quartz ampoules
%Our first crystals were not grown by the
%Bridgman technique, but instead were grown simply by ramping up a
%nominally stoichiometric mixture in an evacuated quartz ampoule to
%$800\,^{\circ}\mathrm{C}$ at $10\,^{\circ}\mathrm{C}$/hr, then cooling
%at a rate of $6\,^{\circ}\mathrm{C}$/hr to
%$500\,^{\circ}\mathrm{C}$, after which the sample was furnace cooled to room temperature. This method
facilitated the formation
of large single crystals, though not with a preferred orientation.  The Cu-doped sample grown following the method described in Reference \onlinecite{HorCuDopingReference}. Powder X-ray diffraction measurements confirmed the phase purity of each sample (see
Fig.~\ref{xrays}). Table~\ref{tab:samples} summarizes the growth methods and properties of all of the samples investigated.

A low speed diamond wheel saw was used to cut rectangular bar shapes from the
large single crystals in our ingots of Bi$_2$Se$_3$. Standard 4-wire
resistivity measurements and 5-wire Hall measurements were performed
using a Quantum Design Physical Properties Measurement System (PPMS), and also with a Keithley current source and
nanovoltmeter in a closed-cycle refrigerator. Magnetoresistance measurements were performed in the
PPMS by using the 4-wire method with silver painted contacts to
measure sample resistance at a temperature of 1.9 K. Shubnikov-de-Haas
oscillations were observed in some, but not all, samples. The carrier
concentrations and mobilities were estimated from the Hall
coefficients and resistivities. The NMR spectra and relaxation rates  were measured using a standard spin echo pulse sequence at varying frequencies  in fields of 9 T and 11.7 T.

\section{Results and Discussion}

\subsection{Charge Transport}

\begin{figure}
\centering
   \subfloat[]{\label{nhallvtemp}
  \includegraphics[width=0.8\linewidth]{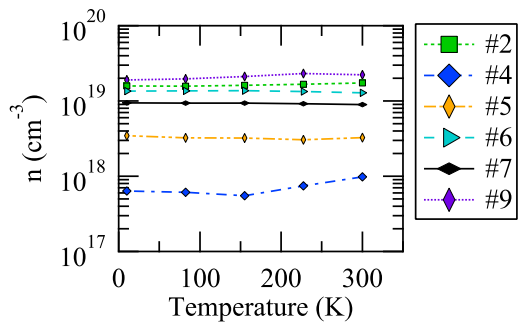}%
   }

  \subfloat[]{\label{resvtemp}
 \includegraphics[width=0.8\linewidth]{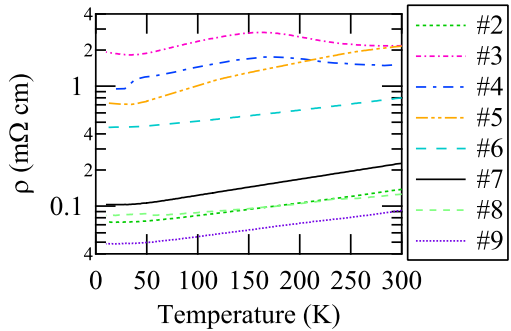}%
  }

  \caption{\label{transport}\protect\subref{nhallvtemp} Carrier concentrations of Bi$_{2}$Se$_{3}$ as determined by Hall
    measurements. The stoichiometric samples (\#2 and \#6 marked as in Fig.~\ref{spectra}) show the highest concentrations, as
    well as \#7 and \#9  (hollow diamonds and slanted bars, respectively) from batch B.  Excess Se (samples marked as in
    Fig.~\ref{spectra}) can lower the concentration by an order of magnitude. \protect\subref{resvtemp} Temperature
    dependence of resistivities of our samples. Sample \#4 results are from the closed-cycle refrigerator; all other
    samples are from the measurements in the PPMS.}

\end{figure}

The electronic properties of our samples were strongly dependent on
growth conditions. The Hall voltage and resistivity of several samples were measured as a function of temperature, and these results are shown in Figs. \ref{nhallvtemp} and \ref{resvtemp}.  The carrier concentration $n$ was estimated from the Hall coefficient, $R_H$ using the formula
$R_H = -1/n q$, where  $q=-e$ is the charge of the
electron carriers (from the Se vacancies). The Hall coefficients of samples from batches B and E have carrier concentrations of order
10$^{19}$ cm$^{-3}$. Samples grown with excess Se (batches C and D)  have lower
carrier concentrations on the order of $10^{18}$ cm$^{-3}$.  Excess Se can reduce the carrier concentration  by reducing the number of Se vacancies, which donate two electrons each. On the other hand crystals grown in excess Se can also have Bi/Se antisite defects, which donate a single electron.\cite{BludskaDefectExperiments} Although our NMR spectra indicate that we introduce some disorder in the samples with excess Se, it is not readily apparent that this disorder arises from antisite defects.

\begin{figure}
\includegraphics[width=\linewidth]{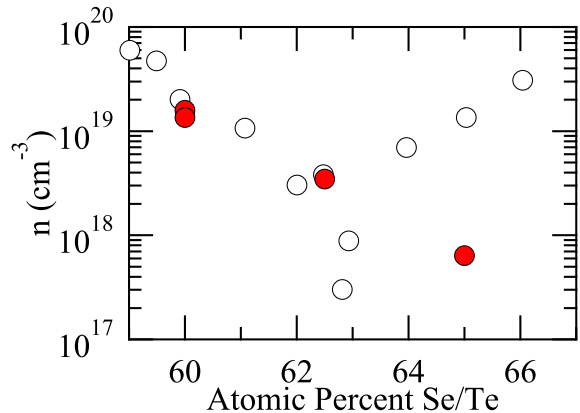}%
 \caption{\label{nHvpSe} Carrier concentration of Bi$_2$Se$_3$ ($\bullet$) and Bi$_2$Te$_3$ ($\circ$) as a function of the atomic percentage of Se/Te in the initial mixture of Bi and Se. Data for the Bi$_2$Te$_3$ reproduced from \onlinecite{carrierTeRef}.}
\end{figure}

\begin{figure}
\includegraphics[width=\linewidth]{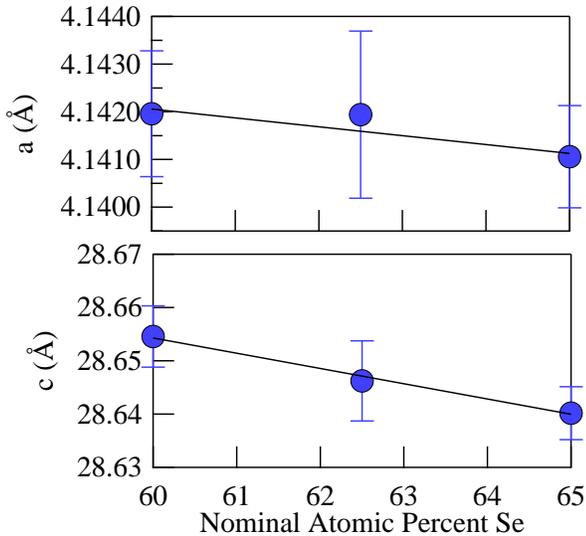}%
 \caption{\label{vegard} Lattice constants $a$ and $c$ as a function of the atomic percentage of Se. Solid lines are best linear fits to the data.}
\end{figure}

Figures ~\ref{nHvpSe} and ~\ref{vegard} display the carrier concentration and lattice constants versus the nominal percentage of Se in these samples.  It is clear that adding Se reduces the number of electrons, consistent with expectation. Furthermore, the monotonic variation of the lattice constants is consistent with Vegard's Law, and indicates that the Se is doping homogeneously in the materials.  Curiously, the behavior of Bi$_2$Se$_3$  contrasts with that of Bi$_2$Te$_3$ (Fig.~\ref{nHvpSe}), in which the carrier concentration first decreases with increasing Te concentration and then  increases.\cite{carrierTeRef} The difference is due to the different types of defects that occur in the two materials: in  stoichiometric Bi$_2$Se$_3$ the dominant defect is  Se vacancies,
so the addition of Se reduces the vacancies and may even add Bi vacancies which act as acceptors. In stoichiometric Bi$_2$Te$_3$ the dominant defect is a substitution of Bi for Te which acts as an acceptor; excess Te causes the dominant defect to be substitution of Te for Bi, a donor.

\begin{figure}
\includegraphics[width=\linewidth]{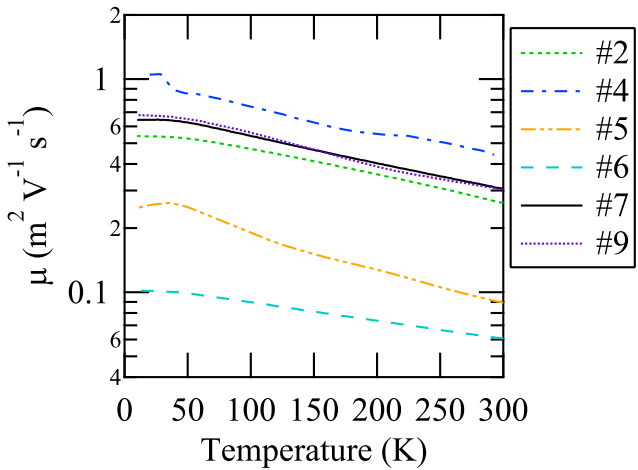}
  \caption{\label{muHvT} The mobilities versus temperature as determined from Hall constant and resistivity measurements.}
\end{figure}

The resistivity, $\rho$, of these samples  revealed metallic
behavior, consistent with a high, temperature independent carrier concentration (Fig.~\ref{resvtemp}) as expected for a degenerate semiconductor. We extract the mobilities, $\mu= 1/\rho n e$ from the resistivity and Hall measurements, as shown in  Fig. ~\ref{muHvT}.  These values are between 1-10 m$^2$/V$\cdot$s and are comparable to those in the literature.\cite{butch2010}

\subsection{Quantum Oscillations}

\begin{figure}
\includegraphics[width=\linewidth]{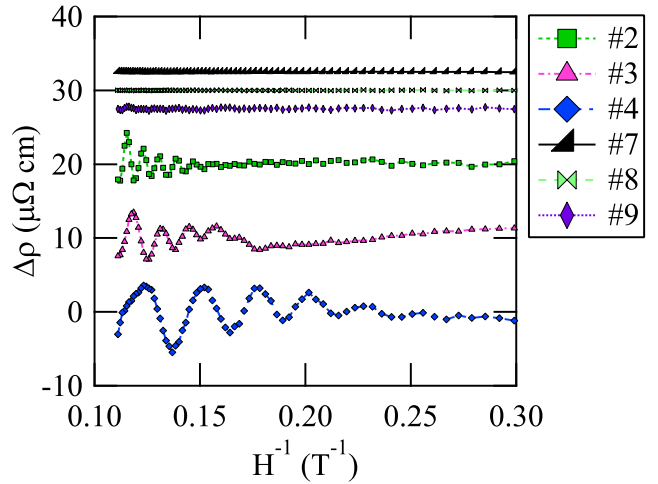}%
   \caption{\label{sdhosc} $\Delta \rho$ versus inverse field, $H^{-1}$ at 1.9 K in Bi$_{2}$Se$_{3}$, where $\Delta \rho$ is the difference between the measured resistivity and a polynomial fit as described in the text. The curves are offset vertically for clarity.  NMR was performed on the samples which showed     oscillations (marked as in Fig.~\ref{spectra}). Other samples from batch B (samples \#7, \#8 and \#9) showed no oscillations.}
\end{figure}

The properties of the doped carriers can be further investigated by magnetoresistance measurements. The resistivity at 1.9 K
was measured up to  a field $H = 9$ T along the $c$-axis and perpendicular to the applied current, and the data were fit to a polynomial of degree five.  The difference, $\Delta \rho$, between the measured values and the polynomial fit display clear Shubnikov-de Haas (SdH) oscillations, as  shown in Fig. \ref{sdhosc}.  The oscillations are described by the Lifshitz-Kosevitch formula:
\begin{equation}
\Delta \rho =  \rho_0 \sin \left[ 2 \pi \left( \frac{F}{H} - \gamma_B \right) \right],
\end{equation}
where $H$ is the magnetic field, $F$ is the SdH frequency related to the Fermi surface area, and $\gamma_B$ is a phase factor related to the Berry phase.\cite{LKReference} The amplitude $\rho_0$ depends on the temperature, the scattering rate,  and the Dingle temperature (defined as $\hbar/2\pi k_{B}\tau$, where $\tau$ is the electron scattering time). From the data we estimate the Fermi wavevectors using the formula:
\begin{equation}
\Delta \left( \frac{1}{H} \right) = \frac{2 \pi e}{\hbar c \mathcal{A}},
\end{equation}
where $\mathcal{A}$ is the maximal cross-sectional area, and assuming the Fermi
surface to be spherical.  Only one sample from batch B showed such
oscillations, with a frequency of about 124 T. Two other samples
grown with excess Se revealed lower SdH frequencies of 72 T
and 35 T (Fig. ~\ref{sdhosc}). These lower frequencies are consistent
with the Fermi surface having a smaller maximal cross-section due to a
lower concentration of electron donors.  These oscillations correspond to Fermi wavevectors $k_F \sim  0.02-0.06$\AA$^{-1}$.  These values are roughly the same order of magnitude as those reported by  ARPES measurements of samples grown by Se self-flux (0.04 \AA$^{-1}$) and traveling
floating zone solvent growth (0.08 \AA$^{-1}$). \cite{young2012} Although we do not have data on the temperature dependence of the SdH amplitudes, we can estimate the upper limit of the effective mass at 1-4 times the bare electron mass based on a fit of the peak-to-peak amplitudes of each cycle versus the inverse field at the midpoints of the cycles, and the measured values of the mobility.

\subsection{Nuclear Magnetic Resonance}

\begin{figure}[hb!]
\includegraphics[width=\linewidth]{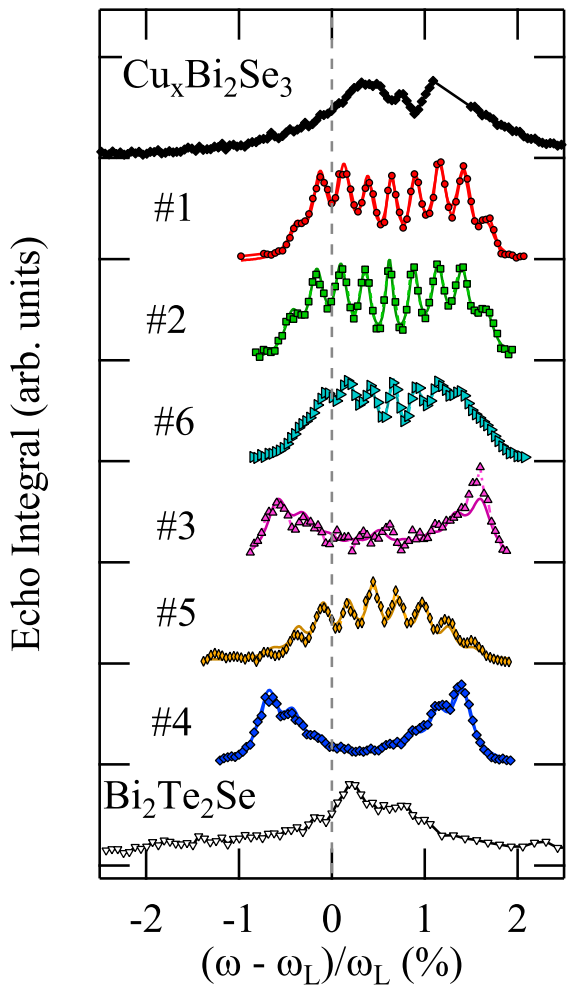}%
   \caption{\label{spectra}$^{209}$Bi NMR spectra of Bi$_2$Se$_3$ single crystals in a field of 9 T with $\mathbf{H}_0\parallel c$ at 10 K (except for \#2 at 20 K). The quadrupolar splitting  appears to be sample-independent, but the Knight shift depends on carrier concentration. Samples \#1 ($\bullet$), \#2 ($\blacksquare$), and \#6 ($\boxplus$) all show a similar spectral
     profile and were all made from stoichiometric mixtures.  Samples
     \#3 ($\blacktriangle$) and \#4 ($+$) have pronounced
     outer peaks, consistent with  observations in samples grown in Se
     self-flux.\cite{young2012} Sample \#5 ($\times$) shows a
     more typical profile. Disorder broadens the quadrupolar satellites and washes out the spectra in the 281 K Bi$_2$Te$_2$Se ($\triangledown$) and Cu$_x$Bi$_2$Se$_3$ ($\blacklozenge$) samples.\cite{ren2010}}
\end{figure}

The Bi NMR spectra were measured  at $10$ K in several different single crystal samples and are summarized in Fig. \ref{spectra}.   The spectra were fit to a sum of nine Lorentzians:
\begin{equation}
S(\omega) = \sum_{n=-4}^4\frac{A_n}{\pi}\frac{\Delta\omega_n}{(\omega - \omega_n)^2 + \Delta\omega_n^2},
\end{equation}
where the frequencies $\omega_n$ are given by Eq. \ref{eqn:freqs}, the linewidths are given by $\Delta\omega_n=\sqrt{(\delta w_M)^2 + n(\delta \omega_Q)^2}$, and $\delta \omega_M$ and $\delta \omega_Q$ are the magnetic and quadrupolar contributions. This equation accounts for both a magnetic and quadrupolar broadening for each transition.\cite{Crocker2011}  The amplitudes $A_n$ were allowed to float, although in principle they are coupled and related to the spin echo decay rate.\cite{young2012}  The fitted values of
$K$, $\delta \omega_M$, $\delta \omega_Q$ and $\nu_{cc}$ are reported in Table \ref{tab:samples}. For all but Bi$_2$Te$_2$Se, the nine satellites of the $^{209}$Bi are distinguishable.  {The EFG splitting $\nu_{cc}\sim 150-170$ kHz} is somewhat smaller than previous experiments on other bismuth compounds.\cite{williams1966,lim1992} For some of the spectra, the intensity of the satellites follow an atypical distribution in which the central resonances are suppressed relative to the satellites. This effect has been observed previously in Bi$_2$Se$_3$ and explained in terms of a spin-spin decoherence rate that depends on the particular nuclear transition.\cite{young2012}  Not all of the spectra exhibit this unusual distribution, however.  In particular, one of the samples grown with excess Se shows a more typical spectral
profile.  This effect probably reflects a longer $T_2$ relaxation time for these samples. Aside from sample \#5, this feature appears to be correlated with the carrier concentration and is more pronounced at lower $n$.

\begin{figure}
\includegraphics[width=0.9\linewidth]{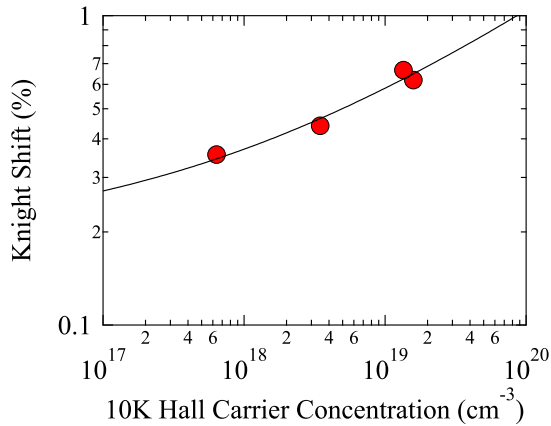}%
   \caption{\label{KvsnH} Knight shift, $K$, versus carrier concentration, $n$. The positive correlation is consistent
     with a hyperfine coupling to electrons. The solid line is a fit as described in the text.}
\end{figure}

Figure \ref{KvsnH} shows the Knight shift, $K$, versus the carrier concentration, $n$, as determined from the Hall constant. The Knight shift arises because spins of the electron carriers couple to the nuclei via the hyperfine interaction.  In order to estimate the magnitude of this coupling, we fit the data to the the expression $K = K_0+A\chi_m$, where the molar susceptibility is given by the Pauli expression:
\begin{equation}
\chi_m = V_m\mu_B^2\frac{m^*}{\hbar^2\pi^2}(3\pi^2 n)^{1/3}.
\end{equation}
Here $n$ is the carrier concentration as determined by the Hall constant, $m^*=0.19m_0$ is the effective mass as determinted by ARPES measurements ($m_0$ is the bare electron mass), $V_m = 258$ cm$^3$/mol is the molar volume, and $A$ is the hyperfine coupling. \cite{ARPESmeasurement}
This expression fits the data well (solid line in Fig. \ref{KvsnH}) with fit parameters $K_0 = 0.19\pm 0.06$\%, and $A = 9.7\pm 1.6$ MOe/$\mu_B$.  In units of energy, $^{209}\gamma\hbar A\mu_B = 27\mu$eV; or alternatively a hyperfine field of  $B_{hf}\sim 735$ G experienced by the electrons.  This value is larger than typical transferred hyperfine couplings, but is comparable to other semiconductors with on-site Fermi contact couplings.\cite{Curro2006a,HyperfineSemiconductors}  It is likely that the hyperfine interaction for the surface states would be similar to that of the bulk. The large value of the hyperfine coupling as well as the high natural abundance of $^{209}$Bi may limit the coherence time of any spin polarized currents, but it is unclear how the spin lattice relaxation process would be modified because of the protected nature of the surface states.\cite{ElectronNuclearInteractionsSemiconductors}

Using this hyperfine constant we can estimate the carrier concentration of the Cu doped sample at $n\approx 2.5\times10^{19}$ cm$^{-3}$ for our sample with a nominal Cu content of $x = 0.12$.   The carrier concentration remains an order of magnitude lower than that of superconducting samples, and probably explains the absence of superconductivity in these samples.\cite{HorCuDopingReference}  Based on the  spectrum seen in Fig. \ref{spectra} it is clear that there is broad distribution of local EFGs that affect the quadrupolar satellites.  This disorder may result from either intercalation of Cu between Bi$_2$Se$_3$ layers or via Cu/Bi substitution.

\begin{figure}
\includegraphics[width=\linewidth]{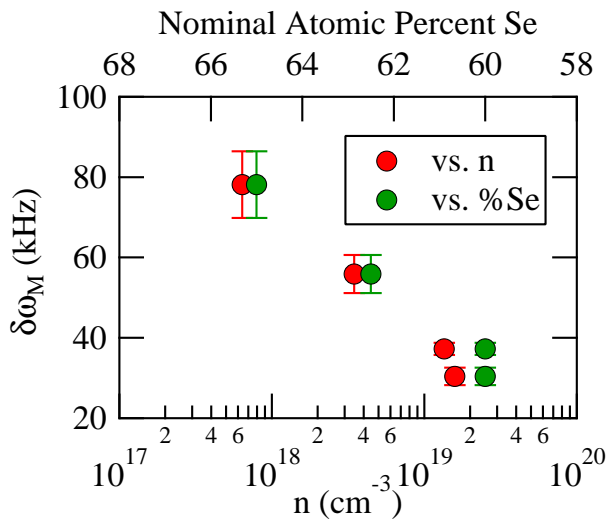}%
  \caption{\label{dKvnH} The magnetic contribution to the linewidth, $\delta \omega_M$ versus carrier concentration (lower axis) and versus Se concentration (upper) axis. }
\end{figure}

Interestingly, there appears to be a negative correlation between the carrier concentration and the magnetic contribution to the linewidth, $\delta \omega_M,$ as shown in Figure~\ref{dKvnH}.
The intrinsic linewidth of the Bi should be determined by the second moment of the nuclear dipole-dipole interaction, which is on the order of 500 Hz, two orders of magnitude lower than our observations.\cite{CPSbook} Defects such as impurities or vacancies in the crystal can
increase this linewidth by creating a local variation in the chemical shift, $K_0$, as well as giving rise to local magnetic fields from localized moments. The fact that that the atypical lineshapes appears to correlate with carrier concentration further suggests the presence of magnetic impurities, since fluctuating magnetic fields enhance the spin decoherence rate, $T_2^{-1}$, and suppress the signal intensity of the central lines. Although excess Se would be expected to narrow the lines by reducing the Se vacancies, in our experiment adding Se seems to broaden the
spectra. This result suggests that excess Se may be introducing lattice defects
other than Se site vacancies, which may be giving rise to localized unpaired electrons. Excess Se may occupy positions normally occupied by Bi as discussed previously, or it may cause Bi vacancies to form which act as acceptors, or it may occupy interstitial positions in the lattice.\cite{SeDoping}
On the other hand, the broadening of the linewidth for low $n$ may reflect an inhomogeneous distribution of local carrier concentrations on a macroscopic scale. This inhomogeneity may be more pronounced at small $n$; however this scenario does not explain the atypical lineshapes.

\begin{figure}
\includegraphics[width=\linewidth]{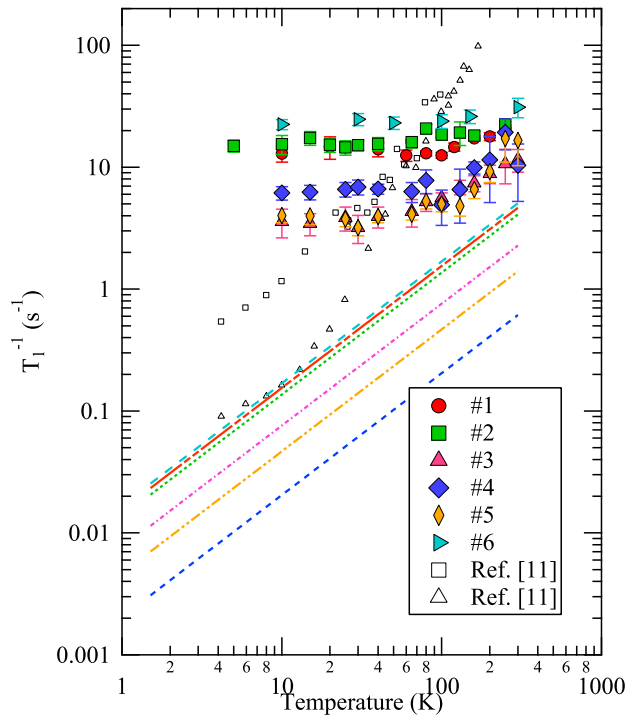}%
   \caption{\label{t1vtemp}Spin-lattice relaxation rates of the $^{209}$Bi versus temperature for various samples with different carrier concentrations.
   The open data points are reproduced from Ref.\ \onlinecite{young2012}, and exhibit a strong temperature dependence.  The solid lines are predictions based on the Korringa formula as described in the text.}
\end{figure}

The \slrrtext\ was measured at the central transition of several of the samples, and the magnetization recovery was fit to the standard expression for spin $9/2$ nuclei:
\begin{eqnarray}
\label{eqn:recovery}
\nonumber M(t) &= M_0\left[1-2f\left(\frac{7938}{12155} e^{-45t/T_1}+\frac{1568}{7293}e^{-28t/T_1}+\right.\right.\\
&{\left.\left.\frac{6}{65}e^{-15t/T_1}+\frac{24}{715} e^{-6t/T_1}+\frac{1}{165} e^{t/T_1}\right)\right]},
\end{eqnarray}
where the equilibrium magnetization, $M_0$, the inversion fraction, $f$, and $T_1$ are fitting parameters.\cite{Narathrecovery}
The results are shown in Fig. \ref{t1vtemp}.  Although our observations are similar in magnitude to other published data, we find a much weaker temperature dependence.\cite{young2012}  In principle there are three possible mechanisms for spin lattice relaxation in a nondegenerate semiconductor: (a) Korringa relaxation via the contact hyperfine interaction with the conduction electrons, (b) spin diffusion from localized electron spins, and (c) quadrupolar relaxation via phonons. We first consider case (a), in which case the \slrrtext should be given by:
\begin{equation}
T_1^{-1} = (K-K_0)^2 T/\kappa
\end{equation}
where the Korringa constant is given by $\kappa = g^2 \mu_B^2/4\pi k_B \gamma^2$ and {$g_{\perp} = 23$ is the  $g$-factor for the electron carriers in this material for the field perpendicular to the $c$-axis.\cite{Bi2Se3gFactor}  Note that although the field lies in the $c$-direction, the fluctuating hyperfine field driving the spin-flip scattering lies in the plane. The calculated values of \slrr\ are shown as solid lines in Fig. \ref{t1vtemp}, and are about an order of magnitude lower than the measured values.  It is possible that the  discrepancy can be attributed to an overestimate of the orbital shift $K_0$ from the fit to Eq. 6 in Fig. \ref{KvsnH}. In fact,  samples \#3, \#4 and \#5 exhibit linear behavior  above 60 K, as expected for Korringa relaxation. On the other hand, for samples \#1, \#2, and \#6 the relaxation the temperature dependence is sublinear, suggesting that the Korringa mechanism is not the dominant relaxation channel for the samples.}

\begin{figure}
\includegraphics[width=\linewidth]{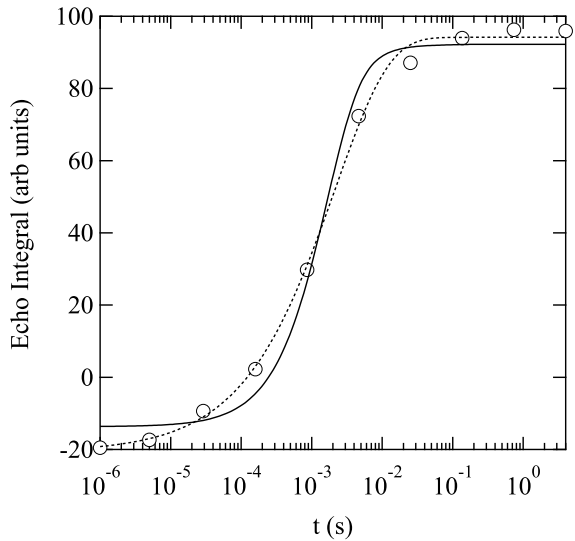}%
   \caption{\label{fig:recovery} Recovery of the Bi echo after inversion recovery at the central line in sample \#2 at 10 K. The solid line is a fit to Eq. \ref{eqn:recovery} and the dotted line is a fit to Eq. \ref{eqn:stretched}.}
\end{figure}

Case (b), spin diffusion from impurities, is a possibility because $^{209}$Bi is 100\% abundant, and Se vacancies/interstitials can bind localized electron spins that are potential sources of scattering/relaxation.  In this case the magnetization recovery typically exhibits a distribution of relaxation rates because spin diffusion drives the relaxation of Bi nuclei with varying distances.  Figure~\ref{fig:recovery} shows the recovery of the magnetization for sample \#2 at 10 K. The solid line is a fit to Eq. \ref{eqn:recovery}, and the dotted line is a fit to the stretched form:
\begin{equation}
\label{eqn:stretched}
M = M_0\left[1-2f e^{-(t/T_1)^\beta}\right],
\end{equation}
where the stretched exponent $\beta$ is a measure of the width of the \slrr distribution.\cite{johnstonstretched}  The data fit better to the stretched exponential form. The temperature dependence of \slrr using Eq. \ref{eqn:stretched} is essentially identical to that of the conventional magnetic relaxation, Eq. \ref{eqn:recovery}, except that \slrr values are about a factor of 50 times larger. However, it is important to note that another explanation for the poor fit to Eq. \ref{eqn:recovery} is that  multiple transitions may have been inverted by the pulse sequence, rather than solely the central transition.  As seen in Fig. \ref{spectra}, the quadrupolar splitting is small ($\sim150$ kHz), and the excitation bandwidth of 167 kHz could have partially excited some of the satellite transitions, especially if one considers that the line width is comparable to the splitting.  As a result the initial conditions of the magnetization recovery would be affected and hence the recovery function (Eq. \ref{eqn:recovery}) would be modified.  Furthermore,  mutual spin flips  between  Bi neighbors, which drives spin diffusion, are likely suppressed  because of the finite quadrupolar splitting.\cite{Curro1998} {The bulk magnetic susceptibility of these samples is diamagnetic, with a clear Curie tail for temperatures below 50 K.  The local moments responsible for this Curie contribution may also be the source of the enhanced magnetic linewidth seen in Fig. 9 as well as the stretched behavior of the relaxation. However there is no evident correlation between $T_1$, $\delta\omega_M$ and the Curie constant, suggesting that either there are extrinsic phases contributing to the susceptibility or the relationship between these quantities is complex.}  Therefore it is unclear which of these two scenarios is the correct explanation for the stretched nature of the magnetization recovery.

Case (c), quadrupolar relaxation, is possible because of the large quadrupolar moment of the Bi.  This mechanism is often difficult to discern, and may occur in combination with magnetic mechanisms.\cite{suterquadrupolarrelaxation}   In this case, the temperature dependence of \slrr is driven by changes in
the lattice and depends on the phonon spectrum. One might expect, then, that \slrr would correlate with the presence of
any lattice imperfections. However, \slrr  does not appear to depend on the electronic mobility, suggesting that this mechanism is not dominant.

\section{Conclusions}

NMR and transport measurements of a series of Bi$_2$Se$_3$ crystals reveal a strong variation in the Knight shift and linewidth as a function of carrier concentration.  By analyzing the Knight shift we determine a hyperfine coupling constant $A\sim 27 \mu$eV.  This coupling leads to a large Knight shift in these samples with high carrier concentrations.  Because this contact hyperfine interaction is determined by the magnitude of the electron wavefunction at the nucleus, it is strongly controlled by large energy scale atomic parameters.  Therefore it should not vary  significantly from one
chalcogenide to another.  Furthermore it should not be significantly different for Bi located on the surface rather than in the bulk.  As a result this electron-nuclear interaction can give rise to scattering of the protected surface state electrons and possibly play a role in dephasing spin polarized currents.\cite{HyperfineDecoherence} The \slrrtext\ we observe is surprisingly large and weakly temperature dependent. Furthermore, the magnetic broadening we observe increases with decreasing carrier concentration.   Our results may suggest the presence of localized spins, possibly associated with Se vacancies or interstitial donor sites that are partially ionized.

\begin{acknowledgments}
We  thank N. apRoberts-Warren for the construction
of the Bridgman furnace, as well as  B-L. Young and D. Yu for stimulating discussions. This work was supported by the National Science
Foundation (Grant No.\ DMR-1005393) and the Eugene Cota-Robles Fellowship.
\end{acknowledgments}

\bibliography{NMR_Bi2Se3_v6}

\end{document}